\documentclass[aoas,MSNbibl,nameyear,seceqn,dvips]{arximspdf}
\usepackage{dcolumn}
\usepackage{graphicx}

\doi{10.1214/11-AOAS497}
\volume{6}
\issue{1}
\pubyear{2012}
\firstpage{284}
\lastpage{303}

\makeatletter

\newcolumntype{d}[1]{D{.}{.}{#1}}

\makeatother

\begin{document}
\begin{frontmatter}

\title{Measuring the vulnerability of the Uruguayan population to vector-borne
diseases via spatially hierarchical factor models}
\runtitle{Measuring the vulnerability}

\begin{aug}
\author[A]{\fnms{Hedibert F.} \snm{Lopes}\corref{}\thanksref{t1}\ead[label=e1]{hlopes@chicagobooth.edu}},
\author[B]{\fnms{Alexandra M.} \snm{Schmidt}\thanksref{t2}\ead[label=e2]{alex@im.ufrj.br}},
\author[C]{\fnms{Esther}~\snm{Salazar}\thanksref{t3}\ead[label=e3]{esther.salazar@duke.edu}},
\author[D]{\fnms{Mariana} \snm{G\'omez}\ead[label=e4]{marianagomezc@higiene.edu.uy}}
\and
\author[E]{\fnms{Marcel} \snm{Achkar}\ead[label=e5]{achkar@fcien.edu.uy}\ead[label=u1,url]{http://www.foo.com}}
\runauthor{H. F. Lopes et al.}
\affiliation{University of Chicago,
Universidade Federal do Rio de Janeiro,
Duke~University, Universidad de la Rep\'ublica and
Instituto Nacional de~Salud P\'{u}blica de Mexico, and
Universidad de la Rep\'ublica}
\address[A]{H. F. Lopes\\
Booth School of Business\\
University of Chicago\\
5807 South Woodlawn Avenue\\
Chicago, Illinois 60637\\
USA\\
\printead{e1}}
\address[B]{A. M. Schmidt\\
Instituto de Matem\'atica\\
Universidade Federal do Rio de Janeiro\hspace*{30.2pt}\\
Caixa Postal 68530\\
CEP.: 21945-970\\
Rio de Janeiro R.J.\\
Brazil\\
\printead{e2}}
\address[C]{E. Salazar\\
Electrical and Computer Engineering\\
Duke University\\
3421 CIEMAS Bldg\\
Box 90291\\
Durham, North Carolina 27708\\
USA\\
\printead{e3}}
\address[D]{M. G\'omez\\
Departamento de Medicina Preventiva y Social\\
Instituto de Higiene\\
Universidad de la Rep\'ublica\\
Av. Alfredo Navarro 3051\\
tercer piso 480 18 67\\
Uruguay\\
\printead{e4}}
\address[E]{M. Achkar\\
Facultad de Ciencias \\
Universidad de la Rep\'ublica \\
Igu\'a 4225 Esq. Mataojo C.P. 11400 \\
Montevideo\\
Uruguay\\
\printead{e5}}
\end{aug}

\thankstext{t1}{Supported by the University of Chicago Booth School of Business.}
\thankstext{t2}{Supported in part by grants from CNPq and FAPERJ.}
\thankstext{t3}{Supported by a Postdoctoral fellowship from FAPERJ.}

% HISTORY:
\received{\smonth{1} \syear{2010}}
\revised{\smonth{6} \syear{2011}}

% ABSTRACT
%
\begin{abstract}
We propose a model-based vulnerability index of the population from
Uruguay to vector-borne diseases. We have available measurements of a
set of variables in the census tract level of the 19 Departmental
capitals of Uruguay. In particular, we propose an index that combines
different sources of information via a set of micro-environmental
indicators and geographical location in the country. Our index is based
on a new class of spatially hierarchical factor models that explicitly
account for the different levels of hierarchy in the country, such as
census tracts within the city level, and cities in the country level.
We compare our approach with that obtained when data are aggregated in
the city level. We show that our proposal outperforms current and
standard approaches, which fail to properly account for discrepancies
in the region sizes, for example, number of census tracts. We also
show that data aggregation can seriously affect the estimation of the
cities vulnerability rankings under benchmark models.
\end{abstract}

% KEYWORDS
%
\begin{keyword}
\kwd{Areal data}
\kwd{Bayesian inference}
\kwd{model comparison}
\kwd{spatial interpolation}
\kwd{spatial smoothing}.
\end{keyword}

\end{frontmatter}

%s1 ###
\section{Vulnerability assessment}
\label{secvuln}
Vulnerability can be defined by a set of
characteristics of a person (or a group of people), which determines
her (or their) ability to anticipate, survive, resist and recover from the
impact of a~dangerous situation
[\citet{blaikiecannondaviswisner1994}]. In addition, \citet{clark2000} mention
that ``questions about vulnerability of social and ecological
systems are emerging as a central focus of policy-driven assessments
of global environmental risks in arenas as different as the ongoing
work of the Intergovernmental Panel on Climate Change, the
World Economic Forum, and the World Food Programme.'' They continue
by saying that ``vulnerability is emerging as a multidimensional
concept involving at least (i) exposure: the degree to which a human
group or ecosystem comes into contact with particular stresses; (ii)
sensitivity: the degree to which an exposure unit is affected by
exposure to any set of stresses; and (iii) resilience: the ability
of the exposure unit to resist or recover from the damage associated
with the convergence of multiple stresses.''
\citet{adger2006} provides a recent review on analytical approaches to
vulnerability to environmental changes. \citet{eakin2006} discuss new
insights into the conceptualization of the vulnerability of
social-environmental systems. They argue that a diversity of approaches
to studying vulnerability is necessary in order to address the full
complexity of the concept.

%s1.1 ###
\subsection{Vulnerability to vector-borne diseases}

In this paper we construct a~micro-environmental index that describes
the vulnerability of the
population of Uruguay to vector-borne diseases, both at the city level,
as well as their census-tracts.
We measure vulnerability by combining the information from a set of
indicators that capture the average social profile of the population,
and the average
environmental condition experienced by households, in a~given census
tract of a city. The very nature of our spatial model (see Section~\ref{secshfm}) allows the assessment of the vulnerability index for the
19 cities in the study (see Section \ref{subsecdata}), as well as other
Uruguayan
cities not included in the study.

In general, approaches that characterize the vulnerability
of human population to vector-borne diseases present problems and
limitations. Some approaches consider poverty as a
determinant indicator, while others consider
climate conditions of key importance to measure vulnerability
[\citet{hahnriedererfoster2009}]. The approaches that assign
special importance to poverty, do so in a classical sense, that is,
they are related to pointing limitations in the availability of
financial resources [\citet{beltrami2008}]. Unarguably, poverty is an
important component of the quality of a person's life. Nonetheless,
the assumption that poverty can solely define a vulnerability index to
vector-borne diseases is a strong and unrealistic one. \citet{adger2006}
points that many authors [e.g., \citet{sen1981} and \citet{sarewitz2003}]
have argued that vulnerability is not the same as poverty. He goes on
mentioning that a vulnerability measure needs to incorporate well-being
defined broadly.

An alternative
approach is to consider
climate characteristics at global, regional and local scales.
These refer to ecological conditions suitable for the development of
the vector, focusing the analysis on the probability of presence of
the vector in the area, and on its ability to
increase its population in some\vadjust{\goodbreak} season of the year
[\citet{lythholbrookbeggs2005}]. These approaches are closely related
to the knowledge of the potential conditions of the development of
the vector under certain conditions, for example, climate variability,
changes in land use and environmental changes, rather than people's
vulnerability to the presence of the vector
[\citet{lieshoutkovatslivermoremartens2004}].

There are in the literature different approaches to assess vulnerability.
For example, \citet{cutter2003} make use of socioeconomic and
demographic data to construct an index of social vulnerability to
environmental hazards (denoted SoVI) for the United States.
Their index is based on a linear combination of factors obtained
through the use of 42 variables observed for all 3,141 U.S. counties.
\citet{schmidtlein2008} analyze the SoVI of \citet{cutter2003}
and discuss its sensitivity to the geographic context under which the
analysis is performed.
\citet{rygel2006} propose a composite index of vulnerability by using
principal components to
aggregate vulnerability indicators. They investigate the vulnerability
of an important United States
metropolitan region to contemporary storm surges and to storm surges
associated with sea-level rise.
More recently, \citet{reid2009} proposed a vulnerability index to heat
waves in the United States. They had available a set of variables for
each of the 39,794 census tracts in the U.S.;
making use of standard factor analysis, they showed that 4 factors
explained more than 75\% of the total observed variance. Their
resultant index is based on a linear combination of these 4 factors.
All the references mentioned above make use of variables which are
observed at some spatial scale. However, none of the analysis above
make use of the information that these variables might be spatially
correlated. We propose to take this information into account when
building such indices.

More specifically, we construct a model-based vulnerability index
by assuming that the set of indicators observed at the census tract
level of a~city
can be probabilistically described by a spatially structured
hierarchical factor model (more details in Section \ref{secshfm}).
The resulting factor is further decomposed into the sum of global and local
effects, which will in turn capture, respectively, the spatial
association across the
cities of the country and within the census tracts of a city. Our
vulnerability index, therefore, takes into account
the covariance among the indicators as well as the covariance across
locations at different spatial scales (point referenced and areal data
level). Moreover, our model-based approach enables the prediction of
the index for unobserved cities and provides guidance to policymakers
regarding vulnerability ranking across the regions.\vspace*{-3pt}

%s1.2 ###
\subsection{Data description}
\label{subsecdata}
\begin{figure}

\includegraphics{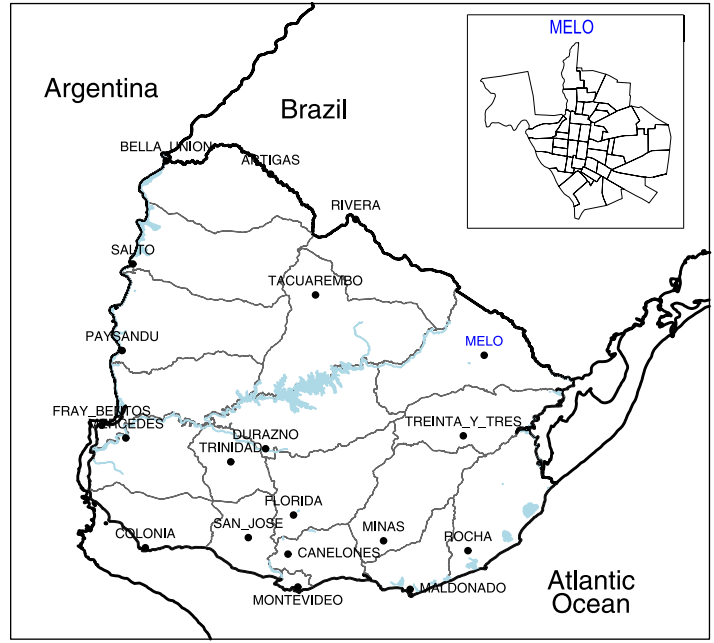}

\caption{Map of Uruguay with department boundaries and capitals
together with the census tracts of Melo.}
\label{Map}
\vspace*{-3pt}
\end{figure}

The Uruguayan territory covers 176,215 km$^2$ with more than 3.4
millions habitants
(roughly the size and the population of the state of Oklahoma), around
50\% of which live in the country's capital, Montevideo.
Uruguay is divided into departments (somewhat similar to states in the
U.S.) with limited local self-government.
Figure \ref{Map} shows the map of Uruguay, its\vadjust{\goodbreak} $I=19$ departments and
their respective capitals.
It also shows the census tracts of Melo (capital of Cerro Lago), in
order to illustrate the spatial information within the city.
Apart form Bella Uni\'on and Montevideo, with 11 and 1,031 census
tracts, respectively, the number of census tracts per capital
varies roughly between 20 and 40 (specific numbers for the other
seventeen capitals appear in Figure \ref{ranking}).

%
%f1 ###
%

A total of $p=11$ indicators are observed at the census tract level for
all nineteen departmental capitals of Uruguay.
These indicators represent the most complete and reliable data set
available in Uruguay and were collected during the most recent
census in 1996 (\textit{Censo Nacional de Poblaci\'on Hogares~y Viviendas
de 1996}).

All the available variables are observed in the census tract level of
the \mbox{$I=19$} Departmental capitals.
They have been standardized to represent percentages, averages,
densities, etc. Broadly,
%
%t1 ###
%
\begin{table}
\tabcolsep=5pt
\caption{Description of the $p=11$ variables,
observed in the census tract level of the departmental capitals, to
build the vulnerability index of the population of Uruguay to
vector-borne diseases}\label{tabvariables}
\vspace*{-3pt}
\begin{tabular*}{\tablewidth}{@{\extracolsep{\fill}}lc@{}}
\hline
\textbf{Level of vulnerability} & \multicolumn{1}{c@{}}{\textbf{Variables}} \\ \hline
Personal characteristic& Illiteracy rate (ILL) \\
& Population with access to public health care (PHC)\\
& Male without formal jobs (UQW) \\
[6pt]
Household characteristic & Owed houses (OWH) \\
& Households headed by a woman (WHF) \\
& Households without sewage system (AHS) \\
& Average number of persons per household (APH) \\
& Households with more than two persons per room (OVC)
\\
& Households without access to treated, drinkable water (ADW) \\
& Households with air conditioner (ACO) \\
& Households poorly built (HOQ) \\
\hline
\end{tabular*}
\vspace*{-3pt}
\end{table}
the available indicators can be divided into two groups representing
general assessments of vulnerability:
personal and household conditions. See Table \ref{tabvariables} for details.

%s1.3 ###
\subsection{Vulnerability via spatial factor models}

As vulnerability is related to weighing the influence of many different
indicators
on a person's (or~group's) living pattern, it is not surprising that
the current literature contains
abundant factor and principal component analysis alternatives to build
such an index.\vadjust{\goodbreak}
Factor models and spatial models are, indeed, two successful examples
of the
broader class of hierarchical models that have been experiencing
major attention from the scientific community as well as from
practitioners in areas as diverse as climate/environment,
economics/finance and health/psychology, among many others. Both areas
directly benefited from the accumulated
advances in Bayesian computation over the last two decades [see
\citet{gamelope2006}].
Fully Bayesian treatments of
factor models and spatial models are described, for instance, in
\citet{lopewest2004} and \citet{banecarlgelf2004},
respectively, and their references.

Various versions of spatial factor models have appeared in recent
years. For instance, \citet{wangwall2003} and \citet{hogatche2004}
model mortality rates and material deprivation measurements,
respectively, by first reducing the dimension of the measurement
vectors at each spatial location via standard factor analysis, and then
spatially modeling the resulting common latent factors. For
spatio-temporal problems, \citet{lopesalagame2008}, for instance,
cluster regions by spatially structuring the factor loadings matrix in
a dynamic factor analysis context.

The remainder of the paper is organized as follows.
We propose in Section~\ref{secshfm} a new class of spatial models,
namely, spatial hierarchical factor models (SHFM), that enables the
construction of vulnerability indices
by combining data observed at all census tracts of some cities of a
country, and also avoids
loss of information and distortions due to data aggregation.
Standard unstructured hierarchical factor models result from our
general proposed model
and are considered as benchmark models for comparison purposes.
Therein, we also discuss a model for observations aggregated at the
city level. Our aim is to investigate, for the entertained models, the
effect of aggregation\vadjust{\goodbreak}
when the vulnerability index is used to rank the cities.
Section~\ref{securuguay} summarizes the
steps toward the construction of a micro-environmental index to
vector-borne diseases for
Uruguay's capitals and census tracts. As the inference procedure follows
the Bayesian paradigm, uncertainty of our estimates are naturally
accounted for. Final remarks appear in Section
\ref{secdiscussion}.\vspace*{-2pt}

%s2 ###
\section{Spatially hierarchical factor model (SHFM)}
\label{secshfm}

Let the region under consideration be divided into capitals, each of
which is further divided into census tracts. This hierarchy can be expanded
down to include more refined layers (levels) depending on the study.
For sake of
simplicity and without loss of generality, we proceed with two levels:
capitals and census tracts.
For each one of the $n_i$ census tracts of capital $i$, a
$p$-dimensional vector of
variables (social-economical, environmental, demographical, etc.) is observed,
namely, $y_{ij}=(y_{ij1},\ldots,y_{ijp})'$, for $i=1,\ldots,I$ and
$j=1,\ldots,n_i$.\vspace*{-2pt}

%s2.1 ###
\subsection{Observational level}
\label{subsec1st}
The $p$ region-specific measurements, denoted here by $y_{ij1},\ldots,y_{ijp}$,
are used to construct a single factor $f_{ij}$ (the vulnerability
factor in our study). More
specifically, the observational level of our model is
%
%e2.1 ###
%
\begin{equation}\label{eqfm}
y_{ijk} = \mu_{k} + \beta_k f_{ij} + \sigma_k\varepsilon_{ijk},\qquad
k=1,\ldots,p,
\end{equation}
where $\mu_{k}$ represents the overall grand mean vector for
measurement $k$.
The factor loadings vector $\beta=(1,\beta_2,\ldots,\beta_p)'$ plays an
important role in understanding
the role and the composition of the common factor. Its first element is
set to one in order to ensure likelihood identifiability
[see \citet{lopewest2004}]. The specific factors $\varepsilon_{ijk}$
are standard normally distributed and
independent across capitals, census tracts and measurements.
The impact of the common factor $f_{ij}$ on $y_{ijk}$
can be assessed, as in standard factor analysis, by the proportion of
the variance of $y_{ijk}$
explained by $f_{ij}$, that is,
%
%e2.2 ###
%
\begin{equation}\label{eqvd}
\pi_{ijk} = \biggl(1+\frac{\sigma_k^2}{\beta_k^2\nu_{ij}^2}\biggr)^{-1},
\end{equation}
where $\nu_{ij}^2=\mbox{Var}(f_{ij})$. Assume, for illustration, that
the variance of the common factor for a given census tract $i$ in a
given city $j$ is equal to one. In this case, $\pi_{ijk} = 1/(1+\sigma
_k^2/\beta_k^2)$ and the proportion of the variance of the $k$th
measurement that is explained by the common factor increases when the
component idiosyncratic standard deviation
decreases relative to the absolute value of its loading.\vspace*{-2pt}

%s2.2 ###
\subsection{Modeling the factor $f_{ij}$}
\label{subsec2nd}

Within capital $i$, the vector of common factors $f_i=(f_{i1},\ldots
,f_{in_i})'$ is
decomposed as the sum of two spatially structured components: one that
captures the overall mean of the capital, and the other one captures
the local structure of the index, in the census tracts' level, and also
accounts for possible effects of neighboring census tracts. More
precisely, we assume
%
%e2.3 ###
%
\begin{equation}\label{HFMeq-f}
f_{ij} = \theta_i+{\tilde f}_{ij} + \sqrt{\omega_i} u_{ij},\vadjust{\goodbreak}
\end{equation}
where $\theta_i$ is the common factor for capital $i$, ${\tilde
f}_{ij}$ is the specific factor for census tract $j$ and capital $i$, and
$u_{ij}$ are independent standard normals.
The error term $u_{ij}$ accounts for unanticipated, location specific
idiosyncrasies. Similar to
(\ref{eqvd}), $\mbox{var}(f_{ij})=\nu_{ij}^2 = \eta_i^2 + {\tilde\nu
}_{ij}^2 + \omega_i$, where $\eta_i^2=\mbox{var}(\theta_i)$ and
${\tilde\nu}_{ij}^2=\mbox{var}({\tilde f}_{ij})$. Therefore, the unexplained
proportion of $\nu_{ij}^2$ (due to $u_{ij}$) is given by
%
%e2.4 ###
%
\begin{equation}\label{eqvd1}
{\tilde\pi}_{ij} = \biggl(1+ \frac{\eta_i^2+{\tilde\nu}_{ij}^2}{\omega
_i}\biggr)^{-1}.
\end{equation}
Large values of $\omega_i$ (relative to $\eta_i^2$ and ${\tilde\nu
}_{ij}^2$) lead to ${\tilde\pi}_{ij}$ close to one and indicates small
explanatory power
of the common factor $\theta_i$ and the specific factor ${\tilde
f}_{ij}$ for census tract $j$ and capital $i$.\vspace*{-3pt}

\subsubsection*{Spatial variation within capitals}
As the capitals are divided into census tracts defining irregular subregions,
we model the within capital factors ${\tilde f}_i=({\tilde
f}_{i1},\ldots,{\tilde f}_{in_i})'$, for $i=1,\ldots,I$, by a proper
conditionally autoregressive (CAR) specification
[\citet{suntsutakawaspeckman1999}]:
%
%e2.5 ###
%
\begin{equation}\label{HFMcar}
\tilde{f}_i \sim N(0,\tau_i^2P_i),
\end{equation}
where $P_i=P_i(\phi)=(I_{n_i}+\phi M_i)^{-1}$, $M_i =D_i-W_i$,
with $w_{ijl}$, the $(j,l)$ component of $W_i$, given by
$w_{ijl}=1/d_{jl}$ if $j$ and $l$ are
neighbors (denoted here by $j \sim l$) and zero otherwise, $d_{jl}=
\|s_j - s_l\|$ is the
Euclidean distance between centroids of capitals $j$ and $l$,
$D_i=\mbox{diag}(w_{i1+},\ldots,w_{in_i+})$ and
$w_{ij+}=\sum_{l \sim j} w_{ijl}$. The inverse matrix $P_i^{-1}$ is
diagonally dominant and positive definite [\citet{harv1997}].
The parameter\vspace*{1pt} $\phi$ controls the strength of the association between
the components of ${\tilde f}_i$, with $\phi=0$ implying independence.
Equation (\ref{HFMcar}) approaches the intrinsic autoregressive model
when $\phi$ approaches infinity [\citet{besayork1991},
\citet{besakoop1995}].
When an intrinsic autoregressive prior distribution is assumed, it is
equivalent to
imposing a spatial structure on the parameters.
We decided for letting the data inform whether the spatial correlation
among the census tracts
is present. However, it is known that there is little information about~$\phi$ above.
We actually tried to fit a model using
the reference prior suggested by \citet{ferreiradeoliveira2007}, but
convergence was
not reached. Therefore, we decided for fixing these parameters at some specific
values and used some model comparison to decide which value fits the
data best.
It is worth pointing out that in the analysis performed in Section \ref
{securuguay} we have also fitted a model assuming an intrinsic
autoregressive prior for these parameters and the results did not
differ much.
The variance of $f_{ij}$ from (\ref{eqvd1}) is then given by
${\tilde\nu}_{ij}^2 = \tau^2_i P_{i,jj}$.\vspace*{-3pt}

\subsubsection*{Spatial variation between capitals}
We assume that the $\theta_i$'s are conditionally independent and
Gaussian with common baseline vulnerability factor~$\theta_0$\vadjust{\goodbreak}
and covariance structure driven by the Euclidean distances between the
centroids of the capitals, that is,
%
%e2.6 ###
%
\begin{equation}\label{HFMtheta}
\theta\sim N(1_I \theta_0,\delta^2H(\lambda)),
\end{equation}
where $\theta=(\theta_1,\ldots,\theta_I)$. Although each capital $i$
has its own vulnerability factor, the above model allows
borrowing-strength across neighboring regions. The correlation matrix
$H$ is fully specified by a Mat\'ern correlation function, that is,
$H_{ij}=\rho(\lambda,d_{ij})= 2^{1-\lambda_2}
\Gamma(\lambda_2)^{-1}(d_{ij}/\lambda_1)^{\lambda_2}\mathcal{K}_{\lambda
_2}(d_{ij}/\lambda_1)$,
where $\mathcal{K}_{\lambda_2}$ is the modified Bessel function of the
second kind and of order $\lambda_2$, $\lambda=(\lambda_1,\lambda_2)$
and $d_{ij}=\|s_i-s_j\|$ is the Euclidean distance between the
centroids $s_i$ and $s_j$ of capitals $i$ and $j$, for
$i,j=1,\ldots,I$. In our application we fixed $\lambda_2=1$ since, as
suggested by \citet{whittle1954}, this value should play an important
role in spatial statistics. It is easy to see that
$\eta_i^2=\mbox{var}(\theta_i)=\delta^2$ for all $i=1,\ldots,I$.
Therefore, (\ref{eqvd1}) can be rewritten as
%
%e2.7 ###
%
\begin{equation}\label{eqvd2}
{\tilde\pi}_{ij} = \biggl(1+ \frac{\delta^2+\tau_i^2P_{i,jj}}{\omega
_i}\biggr)^{-1}.
\end{equation}

%s2.3 ###
\subsection{Posterior inference and model selection}

We now assign the joint prior distribution of the hyperparameters. Here
$W \sim N(a,b)$ means that $W$ is normally distributed with mean $a$
and variance $b$, and $Q \sim \operatorname{IG}(c,d)$
means that $Q$ follows an inverse gamma
distribution whose probability density function evaluate at $q$ is
proportional to $q^{-(c+1)}\exp(-d/q)$.

The joint prior distribution for the remaining parameters is the
product of the following independent marginals:
$\mu=(\mu_1,\ldots,\mu_p)' \sim N(\mu_{0},C_{\mu})$, $\beta_k \sim
N(\beta_0,C_0)$, for $k=2,\ldots,p$, $\sigma_j^2 \sim \operatorname{IG}(a_j,b_j)$ for
$j=1,\ldots,p$, $\omega_i \sim \operatorname{IG}(g_i,h_i)$, $\tau^2_i \sim
\operatorname{IG}(c_i,d_i)$, for $i=1,\ldots,I$, $\theta_0 \sim N(t_0,V_0)$,
$p(\delta^2) \propto1/\delta^2$ and $\lambda_1 \sim \operatorname{IG}(2,h)$, where
$h=d_{\max}/(-2\log(0.05))$ and $d_{\max}$ is the maximum distance
between locations [see \citet{schmgelf2003}, \citet{banecarlgelf2004} for
more details]. Let $\Theta$ be the parameter vector comprising all the
unknowns in the model. The posterior distribution of $\Theta$ is
proportional to
\begin{eqnarray*}
p(\Theta|y) &\propto& \prod_{i=1}^I \Biggl\{\prod_{j=1}^{n_i}
p(y_{ij}|\mu,\beta,f_{ij},\Sigma)\Biggr\}
p(f_i|\theta_i,{\tilde f}_i,\omega_i)
p({\tilde f}_i|\tau_i^2,\phi_i)p(\omega_i)p(\tau^2_i)\\
&&{}\times
\Biggl\{\prod_{k=2}^p p(\beta_k)p(\sigma_k^2)\Biggr\}p(\mu)
p(\sigma_1^2)p(\theta|\theta_0,\delta^2,\lambda_1)
p(\theta_0)p(\delta^2)p(\lambda_1).
\end{eqnarray*}

Closed form posterior inference is infeasible and inference for model
parameters is facilitated by a customized Markov
chain Monte Carlo (MCMC) scheme that combines Gibbs and Metropolis--Hastings
steps. Further details about prior selection and posterior inference
for our SHFM appear in the supplementary material of
\citet{lope2011}.

Model comparison is based on
the deviance information criterion of Spiegelhalter et~al.
(\citeyear{spiebest2002}),
the expected posterior deviation of \citet{gelfghos1998}, and\vadjust{\goodbreak}
the scoring rules of \citet{gneibala2007}. We also compute mean square and
mean absolute errors. Further details about these model selection
criteria appear in the supplementary material of
\citet{lope2011}.\vspace*{-3pt}

%s2.4 ###
\subsection{Vulnerability index at unobserved cities}
It is worth noting that~$\theta$~can contain (potentially several)
unmeasured cities. More specifically, if $\theta\,{=}\,(\theta_g',\allowbreak\theta
_u')'$ with
$\theta_u$ the vector of vulnerability for unmeasured cities, then,
from~(\ref{HFMtheta}), it can be shown that $(\theta_u|\theta
_g)$ is also normally distributed (conditionally on the
hyperparameters) and posterior
inference is directly available. More specifically, if $H_{gg}$, $H_{ug}$
and $H_{uu}$ define the corresponding partition of~$H$, then
the prior mean and prior variance of $(\theta_u|\theta_g)$ are
$1_I\theta_0+H_{ug}H_{gg}^{-1}(\theta_g-1_I\theta_0)$
and
$\delta^2(H_{uu}-H_{ug}H_{gg}^{-1}H_{gu})$, respectively.
See Section \ref{securuguay} for more details and Figure \ref{krig-HFM}
for posterior inference for $\theta_g$ or $\theta_u$
for the Uruguayan study.\vspace*{-2pt}

%s2.5 ###
\subsection{Related factor models}\vspace*{-2pt}
\label{subsecdisc}
\subsubsection*{Unstructured model}
We also fit an unstructured hierarchical factor mo\-del (UHFM) to the
Uruguayan data.
The UHFM is a particular case of our SHFM without accounting for the
spatial dependence. This is done by setting $\tilde{f}_i=0$, $i=1,\ldots
,I$, in (\ref{HFMeq-f}), and assuming $H(\lambda)=I_I$. More
specifically,
\begin{eqnarray*}
y_{ij} &=& \mu+\beta f_{ij}+ \varepsilon_{ij},\\[-2pt]
f_i &=& 1_{n_i} \theta_i + \sqrt{\omega_i} u_i,\\[-2pt]
\theta&\sim& N(1_I\theta_0,\delta^2 I_I)
\end{eqnarray*}
for $\mu$, $f_i$, $\omega_i$ ($i=1,\ldots,I)$, $\theta_0$ and
$\delta^2$ as previously defined. Further details about posterior
inference for this UHFM appear in the supplementary mate\-rial of
\citet{lope2011}. In Section \ref{securuguay} we compare our SHFM to
the UHFM.\vspace*{-3pt}

\subsubsection*{Models for aggregated data}
As information for all census tracts of Uru\-guay is unavailable, a
standard way to proceed would be to build an index based on the
observed mean (across census tracts) of each of the variables at the
city level.
More specifically, for capitals $i=1,\ldots,I$, let $\overline
{y}_i=(\overline{y}_{i1},
\ldots, \overline{y}_{ip})$ be a $p$-dimensional vector of
characteristics such that $\overline{y}_{ik} =
n_i^{-1}\sum_{j=1}^{n_i}y_{ijk}$ $(k=1,\ldots,p)$.
In Section~\ref{securuguay} we compare our SHFM to
when the data are aggregated. For this, we propose
an aggregated spatial factor model (ASFM) for which we assume
%
%e2.8 ###
%
\begin{eqnarray} \label{eqmodelagg}
\overline{y}_i &\sim& N(\mu+ \beta f_i, \Sigma),\nonumber\\[-9.5pt]\\[-9.5pt]
f &\sim& N(1_I \theta_0, \delta^2 H(\lambda))\nonumber
\end{eqnarray}
for $\mu$, $\beta$, $\Sigma$, $\theta_0$, $\delta$ and $H(\lambda)$ as
previously defined, and here $f=(f_1,\ldots,f_I)'$.

Assuming a vector of observations $\overline{y}=(\overline{y}_1,\ldots
,\overline{y}_I)$, the likelihood function is given by
%
%e2.9 ###
%
\begin{equation} \label{eqlikagg}
f(\overline{y} \mid\mu, \beta, f, \Sigma) = \prod_{i=1}^I p(\overline
{y}_{i}|\mu,\beta,f_{i},\Sigma),\vadjust{\goodbreak}
\end{equation}
which is based only on the observations at the $I$ cities.
Unlike our HSFM, the ASFM's vulnerability index is given by the
component $f_i$.
We also consider the simpler case where the components of $f$ are
spatially independent, that is, a simple
aggregated factor model (AFM) for which the matrix $H(\lambda)$ is an
identity matrix of dimension $I$.
A major drawback of these aggregated models is that they do not
take into account the fact that the cities have a~different number of
census tracts.

\subsubsection*{Spatial factor models} Our hierarchical
spatial factor model is closely related to \citet{wangwall2003}
and \citet{hogatche2004} spatial factor models. For instance,
Hogan and Tchernis's spatial factor models are used to
construct one-dimensional model-based deprivation indexes for Rhode
Island's census tracts. Their models are special cases of our SHFM
via equations (\ref{eqfm}) and (\ref{HFMcar}). More specifically,
they consider $I=1$ (Rhode Island) and $n_1=228$ for
(\ref{eqfm}) and a variety of spatial structures for $P$
in~(\ref{HFMcar}).

%s3 ###
\section{Building an Uruguayan vulnerability index}
\label{securuguay}

The models discussed in Section \ref{secshfm} are used to construct the
vulnerability index for the Uruguayan data
described in Section \ref{secvuln}. In particular, we investigate the
effect of aggregating the data across the city.
The MCMC algorithm was run for a total of 30,000 iterations with 10,000
burn-in iterations.
For each model, we ran two parallel chains starting at different
initial points of the parametric space.
Posterior inference was based on the last 20,000 iterations, recording
every 5th iteration in order to avoid possible autocorrelations.
We check the convergence of all chains via Brooks and Gelman's
(\citeyear{broogelm1998}) modification of the Gelman--Rubin statistic.

Table \ref{tab-comparison} presents model comparison results based on
the criteria described in the supplementary material of
\citet{lope2011}. We fit the UHFM with $\theta=0$ and with unknown
$\theta$. We fit the SHFM with $\phi=1$, $5$ and $7$, where the greater
the value of $\phi$, the stronger the spatial correlation between the
components of $\tilde{f}_i$. Our SHFM, with $\phi=5$ or $\phi=7$, is
chosen as the best model by all five criteria. In addition, we have
performed residual analysis to investigate the goodness of fit of the
proposed model. We checked whether the standardized residuals follow a
standard normal distribution. Q--Q plots of the residuals (not shown)
for the best fitted spatial hierarchical factor model indicate that the
quantiles from the model are consistent with the normal quantile. In
addition, Table \ref{tab-comparison} shows MSE and MAE measurements
that provide further information about goodness of fit. Hence, the
following results are based on the SHFM with $\phi=5$.

%
%t2 ###
%
\begin{table}
\caption{\textup{Comparing SHFM and UHFM:} comparing the unstructured
hierarchical
factor (UHFM) and spatial hierarchical factor models (SHFM)
for different values of $\phi$.
Best models appear in italic.
DIC: deviance information criterion,
EPD: expected posterior deviation,
CRPS: continuous ranked probability score,
MSE: mean square error and MAE: mean absolute error.
CRPS are in tens of thousands}\label{tab-comparison}
\begin{tabular*}{\tablewidth}{@{\extracolsep{\fill}}lrd{7.1}rrr@{}}
\hline
& \multicolumn{2}{c}{\textbf{UHFM}} & \multicolumn{3}{c@{}}{\textbf{SHFM}}\\
[-4pt]
& \multicolumn{2}{c}{\hrulefill} & \multicolumn{3}{c@{}}{\hrulefill}\\
\textbf{Criterion} & \multicolumn{1}{c}{$\bolds{\theta=0}$} &
\multicolumn{1}{c}{\textbf{Unknown} $\bolds{\theta}$}
& \multicolumn{1}{c}{$\bolds{\phi=1}$} & \multicolumn{1}{c}{$\bolds{\phi=5}$}
& \multicolumn{1}{c@{}}{$\bolds{\phi=7}$} \\
\hline
DIC & $-$21,445.4 & -21\mbox{,}493.3 & $-$21,785.8
& $-$\textit{21},\textit{827.4} & $-$21,827.0 \\
EPD & 2,557.4 & 2\mbox{,}510.9 & 2,453.1 & 2,433.6
& \textit{2},\textit{432.6} \\
CRPS& 1,030.7 & 1\mbox{,}024.2 & 1,014.2 & \textit{1},\textit{010.3}
& 1,010.3 \\
MAE & 2,397.0 & 2\mbox{,}381.8 & 2,374.5 & \textit{2},\textit{367.9}
& 2,369.1 \\
MSE & 1,222.3 & 1\mbox{,}200.1 & 1,177.2 & 1,169.2
& \textit{1},\textit{168.9} \\
\hline
\end{tabular*}
\end{table}

Table \ref{prop-var-capitals} (columns 2--9) presents the percentage of the
variability of each variable (averaged across census tracts) that is
explained by the vulnerability
index [see (\ref{eqvd})]. The rows are ordered from the
largest to the smallest
percentage of the variance,\vadjust{\goodbreak} starting with the first column (ILL), then
moving to the second column (PHC) for ties, and so on.
%
%t3 ###
%
\begin{table}
\caption{\textup{Variance decomposition.}
Columns 2--9: percentage of the variance explained by the vulnerability
index under model SHFM ($\phi=5$) and averaged across census tracts
within a given capital. Percentages are below 10 for OWH, WHF and ACO.
The rightmost column (VAR) is the percentage of the variance explained
by the common factor $\theta_i+\tilde{f}_{ij}$ averaged across census
tracts [equation~(\protect\ref{eqvd1})]}\label{prop-var-capitals}
\begin{tabular*}{\tablewidth}{@{\extracolsep{\fill}}lccccccccc@{}}
\hline
\textbf{Capital} & \textbf{ILL} & \textbf{PHC} & \textbf{OVC}
& \textbf{UQW} & \textbf{AHS} & \textbf{ADW} & \textbf{APH}
& \textbf{HOQ} & \textbf{VAR}\\
\hline
Bella Uni\'on&95&93&90&79&77&70&58&51 & 77\\
Rivera&92&90&86&72&69&61&48&41 & 93 \\
Treinta y Tres&92&90&86&72&69&61&48&41 & 87\\
Melo&92&90&85&70&67&59&46&39 & 93\\
Salto&91&89&84&69&66&57&45&38 & 92\\
Tacuaremb\'o&91&89&84&69&65&57&44&37 & 90\\
Canelones&91&89&84&68&65&57&44&37 & 77\\
Fray Bentos&91&89&84&68&65&57&44&37 & 90\\
Mercedes&91&89&84&68&65&56&44&37 & 91\\
Durazno&91&88&83&68&65&56&43&37 & 90\\
Colonia&91&88&83&68&64&56&43&36 & 91\\
Rocha&91&88&83&68&64&56&43&36 & 91\\
Paysand\'u&91&88&83&67&64&55&43&36 & 95\\
Trinidad&90&88&83&67&64&55&42&36 & 85\\
Florida&90&88&83&67&63&55&42&35 & 90\\
San Jos\'e&90&88&82&66&63&54&41&35 & 90\\
Maldonado&90&87&82&65&62&53&41&34 & 93\\
Minas&89&87&81&64&61&52&39&33 & 90\\
Montevideo&77&73&64&42&39&31&21&17 & 97\\
[4pt]
Uruguay & 90& 88& 83& 67& 64& 56& 43& 36 & 90\\
\hline
\end{tabular*}
\end{table}
The index impact is higher on ILL, PHC and OVC, each one
representing a different socioeconomic characteristic: education,
health care and household structure, respectively.
These results reveal a strong connection between the index and
education and health. Apart from Montevideo, the pattern is relatively
similar across the country, as indicated by the
bottom row of the table. It is interesting to note that the order of
the capitals also obey a North to South decrease in the impact of the
index with a slight northwest--southeast rotation. This North--South
behavior is clear in the results that follow. In addition, the
rightmost column of Table \ref{prop-var-capitals} presents the
percentages of the variances explained by the common factor $\theta
_i+\tilde{f}_{ij}$ averaged across census tracts [see (\ref
{eqvd1})]. For most of the capitals, the explained variability is
around 90\%, indicating the explanatory power of these two measurements.

\begin{figure}

\includegraphics{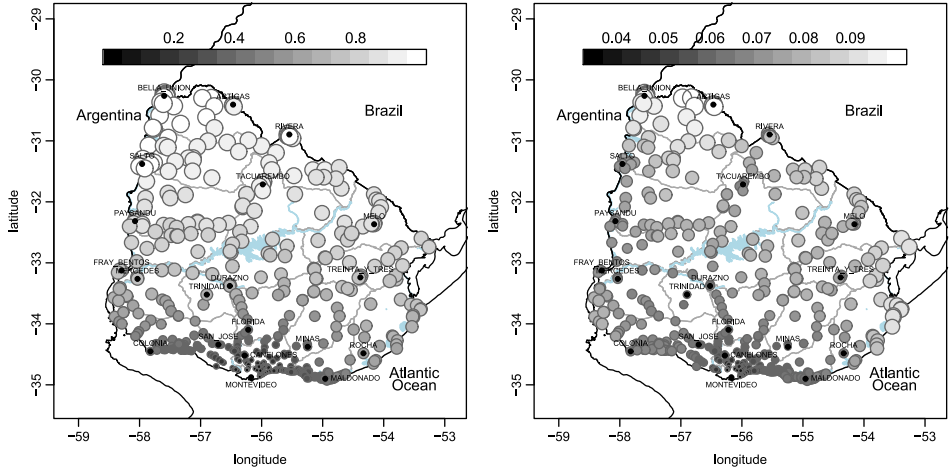}

\caption{Posterior mean of $\theta_i$ and standard deviations (second
column) for observed and unobserved cities under the SHFM when
$\phi=5$.}\label{krig-HFM}
\end{figure}

Figure \ref{krig-HFM} shows the posterior mean and posterior standard
deviations of the component $\theta_i$ of
the vulnerability index for measured and unmeasured cities in Uruguay
under the SHFM with $\phi=5$. We
have standardized the values of $\theta$, and the lower its value,
the better the index.
The index in the country level presents a clear spatial pattern. It
assumes low values in the South of the country and increases
smoothly toward the North--Northwest region of Uruguay. This is in
accordance with the consolidation of urban structures, showing that
$\theta$ is capturing the conditions of the micro-environment of the
population. Although Montevideo concentrates most of the population
of the country, we notice that the index indicates its surrounds as
being the least vulnerable. On the other hand, the North of the
country results in the highest values of $\theta$, corresponding to
the poor conditions of these cities and their suburbs. Also, these
are regions that share a border with Argentina and Brazil, and the
migration within this region is much greater than the improvement
that has been made on basic services for the population. Again, this
component is clearly capturing this characteristic.
Standard errors vary across the region, such that the closer to
monitored locations the lower their values, and are
at most one fifth of the corresponding index.

%
%f2 ###
%

\begin{figure}

\includegraphics{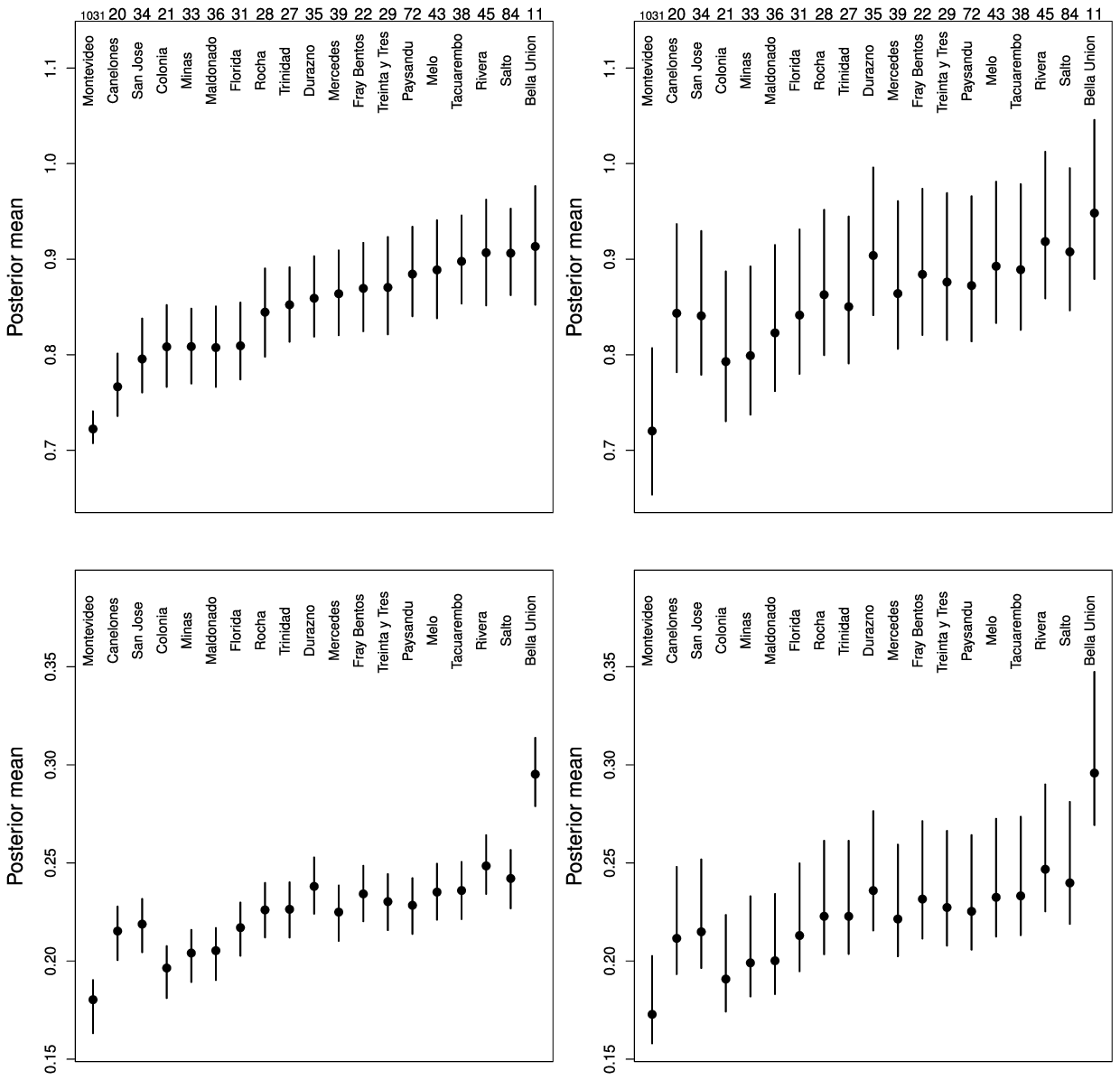}

\caption{Posterior means of the $\theta_i$ (black dots) and 95\% CI
(vertical lines) for each of the 19 capitals under the different
models. \textup{Top row:} SHFM with $\phi=5$ (left) and UHFM (right).
\textup{Bottom row:} ASFM (left) and AFM (right). The quantities on the
top of
the top row boxes are the number of census tracts for each
capital.}\label{rankingtheta}
\end{figure}

Figure \ref{rankingtheta} depicts the effect of either assuming (i) a
spatially structured prior distribution or (ii) an independent prior for
$\theta$, as well as the effect of modeling either (i) the
disaggregated data or (ii) the aggregated data. It is clear that the
range of the posterior distribution of $\theta$ under SHFM is shorter
than that obtained under the assumption that the $\theta$'s are
independent a~priori (model UHFM). Concentrating on the results
when we fit the model for the aggregated data, the spatial model (ASFM)
also results in shorter ranges of the posterior distribution of $\theta
$. However, when comparing SHFM to ASFM, the ranges of the posterior
distribution of $\theta_i$ under
aggregated data (ASFM) do not differ across capitals.

%
%f3 ###
%

This is expected as the model under the aggregated data does \textit{not}
consider the information about the number of census tracts in each
capital. This suggests that the spatial model for the aggregated data
provides conservative estimates of the underlying uncertainty when
estimating $\theta$. Additionally, by ranking the cities based on the
vulnerability index posterior mean, it can be seen that Canelones and
San Jos\'e are at positions 5 and 7, despite their proximity to Montevideo
(under UHFM, ASFM and AFM). Our SHFM corrects this distortion and ranks
these capitals in positions 2 and 3.

%
%f4 ###
%
\begin{figure}

\includegraphics{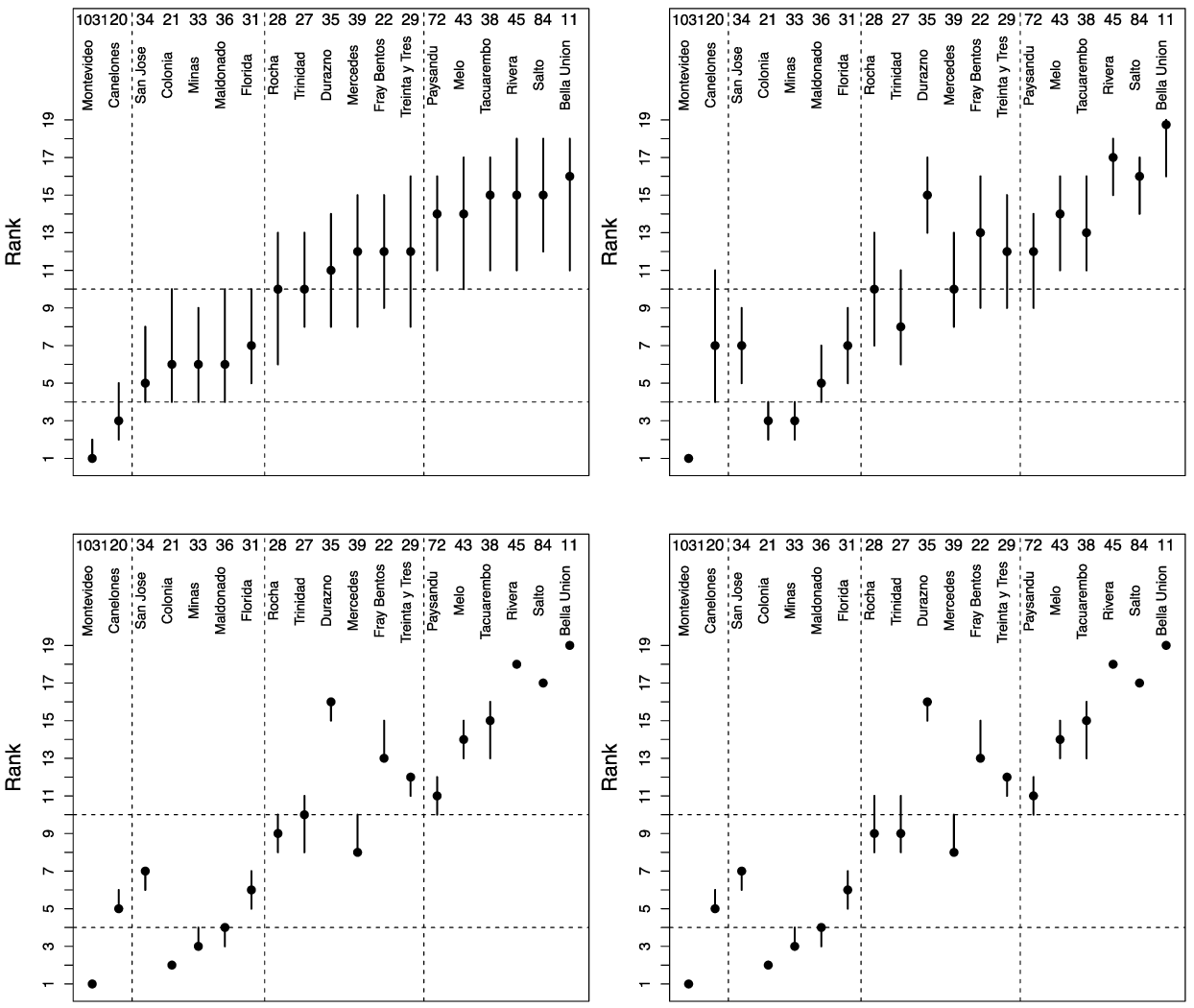}

\caption{Posterior rankings of the capitals. \textup{Top row:} SHFM with
$\phi=5$ (left) and UHFM (right). \textup{Bottom row:} ASFM (left) and AFM
(right). The quantities at the top of each box are the number of
census tracts for each capital.}\label{ranking}
\end{figure}

An important contribution of our modeling strategy is the possibility
of probabilistically ranking the vulnerability across capitals. Figure
\ref{ranking} compares posterior vulnerability rankings based on our
SHFM with $\phi=5$ and the benchmark models, that is, the UHFM, the
ASFM and the AFM. Our SHFM captures the South-to-North spatial
vulnerability increase in Uruguay, as anticipated by the experts. On
the one hand, Montevideo and Canelones are the least vulnerable
capitals, followed closely by San Jos\'e, Colonia, Minas and Maldonado,
all of them located in the South region of the country and all of them
somewhat near Montevideo. On the other hand, Bella Uni\'on, Salto,
Rivera, Tacuaremb\'o, Melo and Paysand\'u are the most vulnerable
capitals, all of them located in the north and northwest regions of the
country. These findings corroborate with our previous findings (see
Figure \ref{krig-HFM}). The UHFM is the model with the closest ranking
pattern, at the capital level, when compared to our SHFM. However, it
suffers from its lack of spatial structure, which leads to different
ranking of the cities, in particular, Canelones, Colonia and Minas.
More critically, the UHFM underestimates the uncertainty associated
with the rankings. Not surprisingly, such behavior is even more marked
under the ASFM and the AFM, where any local structure is distorted or
eliminated by the aggregation of the data. See, for example, the
discussions in \citet{schmidtlein2008}.

\begin{figure}
\begin{tabular}{@{}cc@{}}

\includegraphics{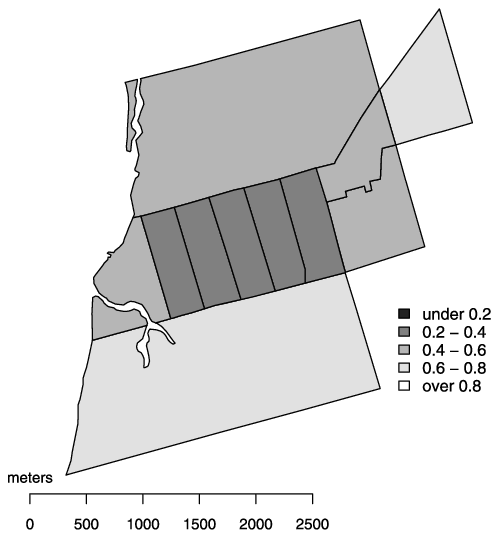}
 & \includegraphics{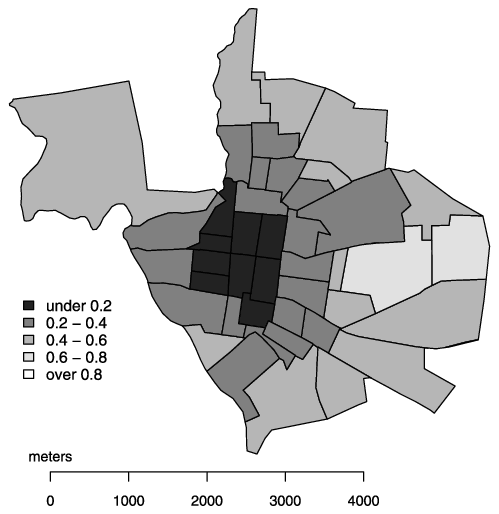}\\
(a) & (b) \\%[4pt]

\includegraphics{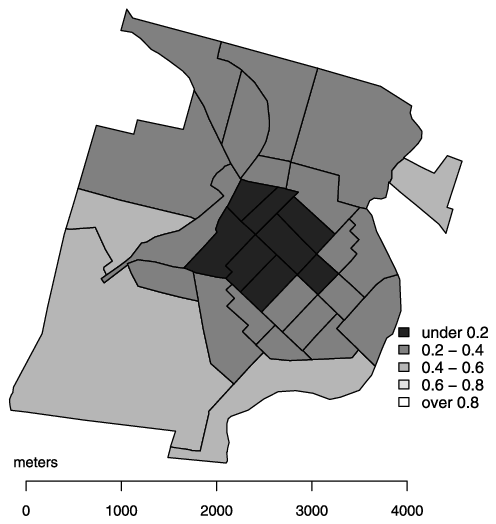}
 & \includegraphics{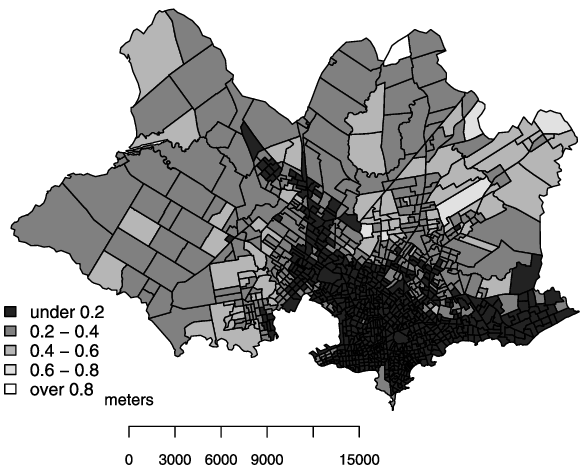}\\
(c)  & (d)
\end{tabular}
\caption{Within-city posterior vulnerability index per census tract of
\textup{(a)} Bella Uni\'on, \textup{(b)} Melo,
\textup{(c)}~Florida and \textup{(d)} Montevideo. Values were standardized
to allow for comparison among the cities. Each map shows
$\kappa_{ij}=(f_{ij}-{\underline f})/({\overline f}-{\underline f})$,
where ${\overline f}=\max_{i,j} f_{ij}$ and ${\underline f}=\min_{i,j}
f_{ij}$.}
%Between-city posterior vulnerability index $\kappa_{ij}=(f_{ij}-{
%and ${\underline f}=\min_{i,j} f_{ij}$.}
\label{mapas-vuln}
\vspace*{-3pt}
\end{figure}

As we propose a factor model for data observed in the census tract
level, following our SHFM, we are able to investigate the components of
the index at each census tract of each city in the sample. Panels of
Figure \ref{mapas-vuln} show the posterior mean of a standardized
version of $f_{ij}$
%
%f5 ###
%
(again under SHFM with $\phi=5$) for each census tract from Bella Uni\'
on, Melo, Florida and Montevideo. Standardization was an artifact to
make the country level effects visually comparable. More precisely, the
standardized within-city posterior vulnerability index is
given by $\kappa_{ij}=(f_{ij}-{\underline f})/({\overline f}-{\underline
f})$, where ${\overline f}=\max_{i,j} f_{ij}$ and ${\underline f}=\min_{i,j} f_{ij}$.
These maps provide evidence of the potentiality of our model in
decomposing the index as the sum of global and local effects.

In panel (a) we have the posterior mean of $\kappa_{ij}$ for the $11$ census
tracts of Bella Uni\'on. The city has lower vulnerability in its
center, where more infrastructure
and more favorable environmental conditions can be found. An
interesting point is that the model
is able to differentiate the two census tracts with more controversial
environmental conditions in the city.
In panel~(b) we have the posterior mean of $\kappa_{ij}$ for the $43$
census tracts of Melo which share the border with Brazil.
The main activity of this conservative region is cattle raising, and
this is concentrated on a small percentage of its population. For this
reason, this is a region with high levels of informal activities and
lack of basic public services, specially in its outskirts. This is what
the estimates of $\kappa_{ij}$ indicate; the central region has values
comparable to the good
conditions that can be found in the South of the country.
Panel~(c), on the other hand, shows the posterior mean of the local
effects for the~$31$ census tracts of Florida, the main city in the area of dairy
production in the country. During the twentieth century, Florida had
a good socioeconomic status. Also its location, in the half plain of
Santa Lucia Chico's river, allowed it to achieve high standards in
environmental terms. With the applications of neo-liberal policies
during the 80s and 90s, small dairy producers lost profitability and
left the sector. This resulted in a migration to the periphery of
the city, which happened much faster than the development of the
necessary urban infrastructure. This is clearly represented by the
``concentric rings'' in panel (c). That is, Florida has good
micro-environmental conditions at the developed city center, and the
levels of these conditions decrease with the increase of the
distance to its center.

Montevideo, the capital of the country, appears in panel (d) with its~$1\mbox{,}031$ census tracts.
The distribution of $\kappa_{ij}$
across the city allows one to discriminate between\vadjust{\goodbreak} very opposing
situations, varying from very low to high values of the parameter,
capturing the local effects of the index. This is clear in the
richest area of the city (Southeast) where there are few census
tracts with high values of the index, representing high
vulnerability. Land in these regions is irregularly owned. Overall,
the levels of $\kappa_{ij}$ in Montevideo are in accordance to what is
anticipated by experts, showing high values (more~vul\-nerability) toward the North--West region which comprises
more rural areas.\vspace*{-3pt}

%s4 ###
\section{Discussion}
\label{secdiscussion}
We proposed a spatially structured factor model to build a
vulnerability index based on measurements
observed at the census tracts level of a country.
In our specific case, we had available observations of $p=11$
indicators at each of the census tracts of the $I=19$ Departmental
capitals of Uruguay.

A key issue in our data set is that the number of census tracts in
Montevideo is much larger than any of the other capitals, and any
factor analysis must take this information into account. To this end,
our proposed model provides an index at each census tract which is
decomposed as the sum of an overall capital effect and a local effect.
In our model the number of census tracts in each capital is naturally
accounted for, as described in (\ref{HFMeq-f}).
Also, our model allows the overall effect and the local effects of a
city to be spatially structured, and independent models are particular
cases of the general structure proposed in Section \ref{secshfm}. Model
comparison can be used to point which model fits the data best. We
entertained among 5 different criteria, and all of them agreed that for
our data set it is better to use a model with a spatially structured
prior for both $\theta_i$ and $\tilde{f}_{i}$.

As inference is performed following the Bayesian paradigm, we are able
to obtain summaries of the posterior distribution of any function of
the parameters.
In particular, our model-based approach provides the estimated ranking
of the cities according to the estimated index under the different
models (SHFM, UHFM, ASFM and AFM).
From the panels in Figure \ref{ranking} it is clear that the
aggregation seriously affects the estimation of the ranking of the
vulnerability of the cities. This is expected, as the likelihood
function does not consider the different number of census tracts among
the cities. Moreover, the spatial structure, anticipated by experts, is
lost when the index is estimated based on the aggregated data.
Our results indicate that in Uruguay this vulnerability index increases
from the South to the North of the country, assuming higher values in
the regions close to the border with Brazil and Argentina.

When the goal is the estimation of an index, similar to the ones we
develo\-ped here, one is advised to
carefully and meticulously understand and explore the data and its
aggregation structure before proposing
any inferential and model selection strategies. Specifically, when the
data comprise spatially referenced observations, it is important to
explore models which allow for spatial dependence. It is also
critically important to acknowledge that aggregating observations might
lead to different, perhaps misleading, results.\vadjust{\goodbreak}

The resultant index is a valuable management tool in public health.
For a country with limited funds such as Uruguay, setting funds
allocation priorities based on solid scientific criteria can be a major
challenge. Our study aimed at providing such a tool. The next step is
to validate our estimated index. This is usually done by performing a
qualitative assessment of the index. For example, \citet{obrien2004}
performed local case studies by visiting highly vulnerable and less
vulnerable districts. They interviewed government officials and also
nongovernmental organizations; household surveys were also carried out.
This is the next step of our research project.

\section*{Acknowledgments}

The authors are grateful to IDRC, Canada, for providing the grant that
made the data acquisition possible. The authors are grateful to two
referees and the Editor whose comments greatly improved the
presentation of the paper.

%suskaldyti doi

\begin{supplement}
\sname{Supplement A}\label{suppA}
\stitle{MCMC scheme and Model Selection\\}
\slink[doi]{10.1214/11-AOAS497SUPPA}
\slink[url]{http://lib.stat.cmu.edu/aoas/497/supplement-A.pdf}
\sdatatype{.pdf}
\sdescription{The full conditional distributions for both the spatially
hierarchical factor model (SHFM) and the unstructured hierarchical
factor model (UHFM) are presented in this supplement. We also provide a
brief overview of the model comparison criteria used in the paper,
namely, (i) expected posterior deviation (EPD), (ii) deviance
information criterion (DIC), (iii) continuous ranked probability score
(CRPS), (iv) mean absolute error (MAE), and (v) mean square error (MSE).}
\end{supplement}

\begin{supplement}
\sname{Supplement B} \label{suppB}
\stitle{\texttt{Ox} Code for SHFM\\}
\slink[doi]{10.1214/11-AOAS497SUPPB}
\slink[url]{http://lib.stat.cmu.edu/aoas/497/Supplement_B.zip}
\sdatatype{.zip}
\sdescription{The folder \texttt{data} contains files
with the 11 socio-economic indicators (the columns of the files)
observed at the census tract level (the rows of the file) for each one
of the 19 Uruguayan capitals (\texttt{montevideo.txt}, for instance, has
1,031 rows and 11 columns). The folder \texttt{neigmat} contains 19 files
with the neighborhood matrices for each one of the 19 capitals after
rearranging the numbering of the census tract using the GMRFLib-library
of \citet{ruefoll2007}. The files \texttt{shfm.ox} and \texttt{functions.ox}
contain the \texttt{Ox} code to perform MCMC-based posterior inference for
our spatially hierarchical factor model (SHFM).}
\end{supplement}

% imsref loaded by lrinkeviciute, 2011-09-09 15:48:25
%
% imsref loaded by lrinkeviciute, 2011-09-12 08:32:39

\printaddresses

\end{document}